\documentclass[letterpaper]{article} 
\usepackage{aaai2026}  
\usepackage{times}  
\usepackage{helvet}  
\usepackage{courier}  
\usepackage[hyphens]{url}  
\usepackage{graphicx} 
\usepackage{natbib}  
\usepackage{caption} 
\usepackage{algorithm}
\usepackage{algorithmic}
\usepackage{multirow} 
\usepackage{multicol} 
\usepackage{booktabs} 
\usepackage{pifont} 
\usepackage{subcaption} 
\usepackage{amsmath} 
\newcommand{\eg}{e.g.}
\usepackage{newfloat}
\usepackage{listings}
\DeclareCaptionStyle{ruled}{labelfont=normalfont,labelsep=colon,strut=off} 
\floatstyle{ruled}
\newfloat{listing}{tb}{lst}{}
\floatname{listing}{Listing}
\newcounter{corrcounter} 
\title{UniMGS: Unifying Mesh and 3D Gaussian Splatting with Single-Pass
Rasterization and Proxy-Based Deformation}
\author{
    Zeyu Xiao\textsuperscript{\rm 1}\equalcontrib,
    Mingyang Sun\textsuperscript{\rm 1}\equalcontrib,
    Yimin Cong\textsuperscript{\rm 1},
    Lintao Wang\textsuperscript{\rm 1},
    Dongliang Kou\textsuperscript{\rm 1},
    Zhenyi Wu\textsuperscript{\rm 1},
    Dingkang Yang\textsuperscript{\rm 1},
    Peng Zhai\textsuperscript{\rm 1},
    Zeyu Wang\textsuperscript{\rm 3, 4}\footnote{These authors are equal corresponding authors.}\setcounter{corrcounter}{\value{footnote}}, 
    Lihua Zhang\textsuperscript{\rm 1, 2}\footnotemark[\value{corrcounter}] 
}
\affiliations{
    \textsuperscript{\rm 1}College of Intelligent Robotics and Advanced Manufacturing, Fudan University, Shanghai, China\\
    \textsuperscript{\rm 2}Fysics Intelligence Technologies Co., Ltd., Shanghai, China\\
    \textsuperscript{\rm 3}The Hong Kong University of Science and Technology (Guangzhou), Guangzhou, China\\
    \textsuperscript{\rm 4}The Hong Kong University of Science and Technology, Hong Kong, China\\

}

\usepackage{bibentry}

\begin{document}

\maketitle

\begin{abstract}
Joint rendering and deformation of mesh and 3D Gaussian Splatting (3DGS) have significant value as both representations offer complementary advantages for graphics applications.
However, due to differences in representation and rendering pipelines, existing studies render meshes and 3DGS separately, making it difficult to accurately handle occlusions and transparency.
Moreover, the deformed 3DGS still suffers from visual artifacts due to the sensitivity to the topology quality of the proxy mesh.
These issues pose serious obstacles to the joint use of 3DGS and meshes, making it difficult to adapt 3DGS to conventional mesh-oriented graphics pipelines.
We propose UniMGS, the first unified framework for rasterizing mesh and 3DGS in a single-pass anti-aliased manner, with a novel binding strategy for 3DGS deformation based on proxy mesh.
Our key insight is to blend the colors of both triangle and Gaussian fragments by anti-aliased $\alpha$-blending in a single pass, achieving visually coherent results with precise handling of occlusion and transparency.
To improve the visual appearance of the deformed 3DGS, our Gaussian-centric binding strategy employs a proxy mesh and spatially associates Gaussians with the mesh faces, significantly reducing rendering artifacts.
With these two components, UniMGS enables the visualization and manipulation of 3D objects represented by mesh or 3DGS within a unified framework, opening up new possibilities in embodied AI, virtual reality, and gaming.
We will release our source code to facilitate future research.
\end{abstract}



\section{Introduction}
Recent advances in radiance representations, especially 3D Gaussian Splatting (3DGS)~\cite{3dgs}, have enabled the efficient modeling of photorealistic 3D scenes.
Adding meshes into a radiance scene for interaction and visualization has shown great value in applications like embodied AI~\cite{xie2025vid2sim} and gaming~\cite{xia2024video2game}.
Consequently, there is an urgent need in the 3D vision community to improve the compatibility between 3DGS and mesh for both rendering and manipulation.
\begin{figure}[t]
    \centering
    \includegraphics[width=0.47\textwidth]{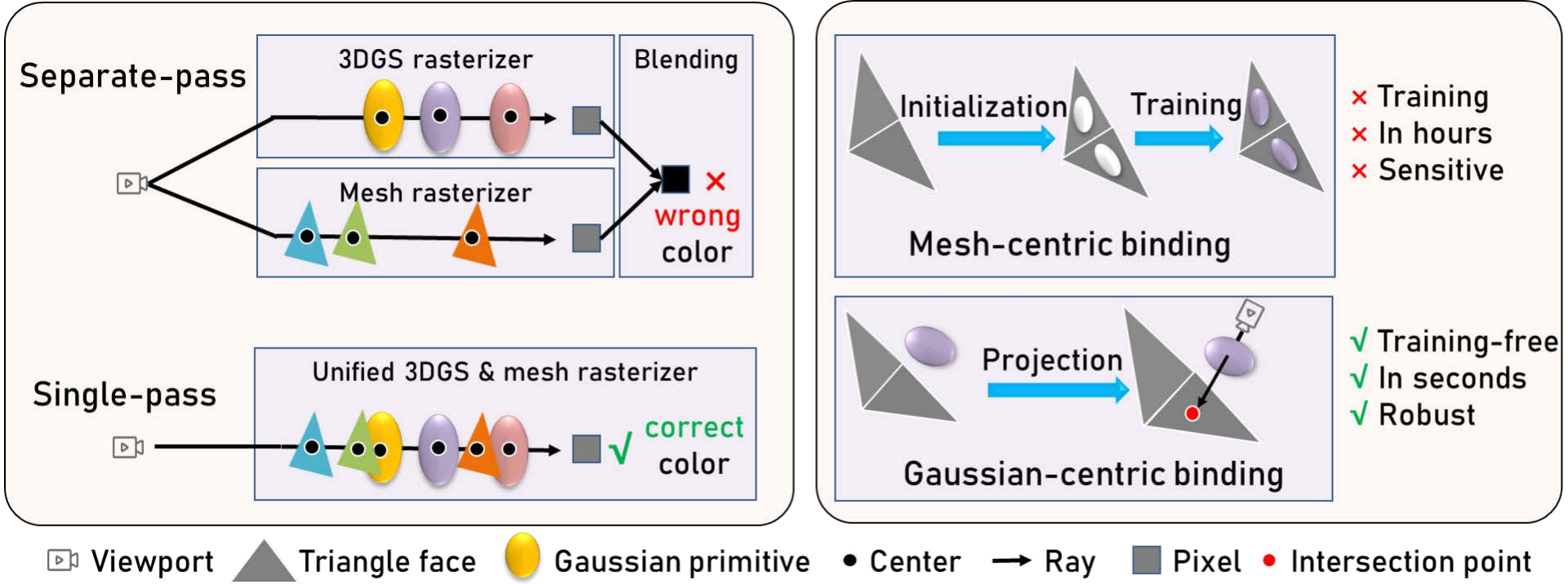}
    \caption{
    Differences between UniMGS and prior work in rasterization and deformation.
    }
    \label{fig:sec0}
\end{figure}

To account for the differences in the rendering between the shape representations, most methods first render meshes and 3DGS separately and then blend the results via visibility and depth sorting, enabling a straightforward separate-pass implementation~\cite{xie2025vid2sim,mega,bridging}.
Since the separate-pass scheme must render each shape individually to calculate visibility, it is inherently unable to handle nested spatial relationships between objects.
Instead, a single-pass method enables the simultaneous processing of both meshes and a radiance representation within a unified rendering pipeline ~\cite{nerfmesh-iccv}, which remains insufficiently explored. 
Although ray tracing has recently allowed for rendering 3DGS and meshes~\cite{raysplats,3dgut}, its run-time performance remains inferior to rasterization due to high computational cost.

In addition, since shapes represented by 3DGS consist of discrete Gaussian kernels with center points, deformation can be achieved by directly manipulating them~\cite{4dgs,physgaussian}.
An alternative approach deforms 3DGS indirectly by attaching it to a proxy and manipulating the proxy, e.g., sparse control points~\cite{scgs,MSGS}, deformation graphs~\cite{arapgs}, cages~\cite{cageGS1,cageGS2}, meshes~\cite{gaussianmesh,games,manigs}. 
As mentioned in GaussianMesh~\cite{gaussianmesh}, paradigms other than those based on mesh proxies lack explicit topological information, leading to misalignment artifacts when handling large deformations.
However, existing approaches that rely on proxy meshes typically adopt a mesh-centric binding strategy, where Gaussians are first initialized on each face and then refined through training with geometric constraints from the proxy mesh.
As they follow an ``initialization-then-training" workflow, even an optimized 3DGS must be retrained to bind with a proxy mesh, which forms a causal dependency.
This design not only reduces usability but also is sensitive to the mesh's topology, leading to artifacts when the mesh contains defects.
Motivated by these observations, we propose \textbf{UniMGS}, a \textbf{Uni}fied framework for \textbf{M}esh and 3D\textbf{GS} integration in single-pass rasterization and proxy-based deformation.
Our core idea consists
of two components: 1) adapting the 3DGS rasterizer for anti-aliased mesh rendering; and 2) adapting mesh-oriented manipulation to 3DGS.
Since both 3DGS and meshes are rasterizable, directly assigning an opacity attribute to textured meshes allows for unified rendering through $\alpha$-blending.
However, the implementation of anti-aliasing differs between the two approaches: for 3DGS, the Elliptical Weighted Average (EWA) filter~\cite{ewa2,3dgs} is applied before $\alpha$-blending. In contrast, for triangle meshes, anti-aliasing is typically performed during $\alpha$-blending and involves a coverage value that depends on the sub-pixel transmittance~\cite{OpenGLbook}.
Therefore, the $\alpha$-blending in 3DGS must be adapted to account for anti-aliasing of meshes.
Our method emerges naturally from this line of reasoning.
We treat the depth-adjacent triangles as a single entity, within which color blending is still performed at the pixel level, while transmittance is independently computed at the sub-pixel level and weighted by coverage to represent pixel-level transmittance.
Outside the entity, each triangle is still considered an individual fragment, since anti-aliasing only applies within this triangle entity.
Therefore, transmittance must be updated among each triangle fragment at the pixel level to contribute to subsequent calculations, such as blending with Gaussians or the background.
This novel design successfully integrates anti-aliased mesh rendering into the 3DGS rasterizer without any degradation, thereby enabling Gaussians and meshes to be rasterized together in a single-pass manner.
To the best of our knowledge, no prior work has explored single-pass rasterization of mesh and 3DGS---a fundamental technology that has been overlooked by the 3D vision community.
To adapt mesh-oriented manipulation to 3DGS, we propose a novel Gaussian deformation representation that enables topology-aware manipulation of 3DGS.
Given the 3DGS of an object and its mesh, which is easily acquired via various reconstruction methods~\cite{gof,neus2}, 
our goal is to link each Gaussian to the mesh faces to allow propagation of the deformation.
To enhance usability and deformation robustness, we advocate a Gaussian-centric perspective to eliminate the causal dependency involved in the mesh-centric ones.
The spatial relationship between Gaussians and mesh faces can be directly utilized for binding.
Thanks to cameras in the training dataset, we propose using ray casting as an accurate and efficient solution.
Specifically, a ray is cast from a camera through the center of a Gaussian; if it intersects a face of the proxy mesh, we record that face.
After traversing all cameras, multiple faces may be obtained. We select the nearest face as the binding target.
During manipulation, deformation is transferred from the mesh to the Gaussian.
We further extend this method to the Bounding Box (BBX) of Gaussians to enhance deformation robustness, as detailed later.
Compared to mesh-centric designs, our Gaussian-centric strategy naturally leverages spatial relationships to associate Gaussians with faces without retraining, thereby achieving higher robustness to mesh imperfections during deformation.

We extensively evaluate \textit{UniMGS} across diverse experiments.
It achieves the first unified rasterization of meshes and 3DGS with faithful transparency, color, and correct occlusion handling.
Compared to existing mesh-centric paradigms, our Gaussian-centric binding strategy delivers superior visual quality of deformation, even with a flawed proxy mesh, significantly outperforming prior methods.
\textit{UniMGS} bridges the gap between 3DGS and mesh while maintaining high flexibility in use.
Both the rasterization and deformation modules independently improve the compatibility between 3DGS and mesh, and their integration further empowers simulations under the hybrid representations.

\section{Related Work}
\subsection{Rendering Methods for Hybrid Representations}
There are two primary paradigms for rendering 3D primitives onto an image plane: ray-based (\eg, ray marching, ray tracing) and primitive-based (rasterization)~\cite{CGBOOK}, both of which are commonly used for meshes.
The Neural Radiance Field (NeRF)~\cite{nerf} adopts ray marching to enable volumetric rendering, while 3DGS~\cite{3dgs} utilizes rasterization for fast rendering.
Despite sharing similar rendering principles, bridging different shape representations within a unified rendering framework remains challenging. 
Most studies render the mesh separately first, then use its depth map to account for occlusion when rendering the radiance representation. 
In mesh-NeRF hybrid rendering, it is commonly assumed that the mesh is opaque, allowing ray marching to terminate early once the mesh surface is hit~\cite{nerfmesh-tvcg,guo2023vmesh,xia2024video2game}.
In mesh-3DGS hybrid rendering, existing studies~\cite{bridging,mega,xie2025vid2sim} first rasterize meshes and then combine the rasterized result with the rendered Gaussians by $\alpha$-blending.
Instead, DMERF~\cite{nerfmesh-iccv} achieves mesh-NeRF coupling in a single-pass manner by allowing rays to alternate between ray tracing and ray marching.
It is evident that single-pass approaches offer substantial advantages over separate-pass methods in accurately calculating transparency and occlusion.
Recent studies~\cite{3dgrt,3dgut,raysplats} have proposed using ray tracing to render 3DGS, enabling seamless integration with meshes. While they improve visual quality, it comes at the cost of nearly halving the rendering speed.



\begin{figure*}[t]
\centering
    \begin{minipage}[c]{0.98\textwidth}
    \centering
    \includegraphics[width=0.98\textwidth]{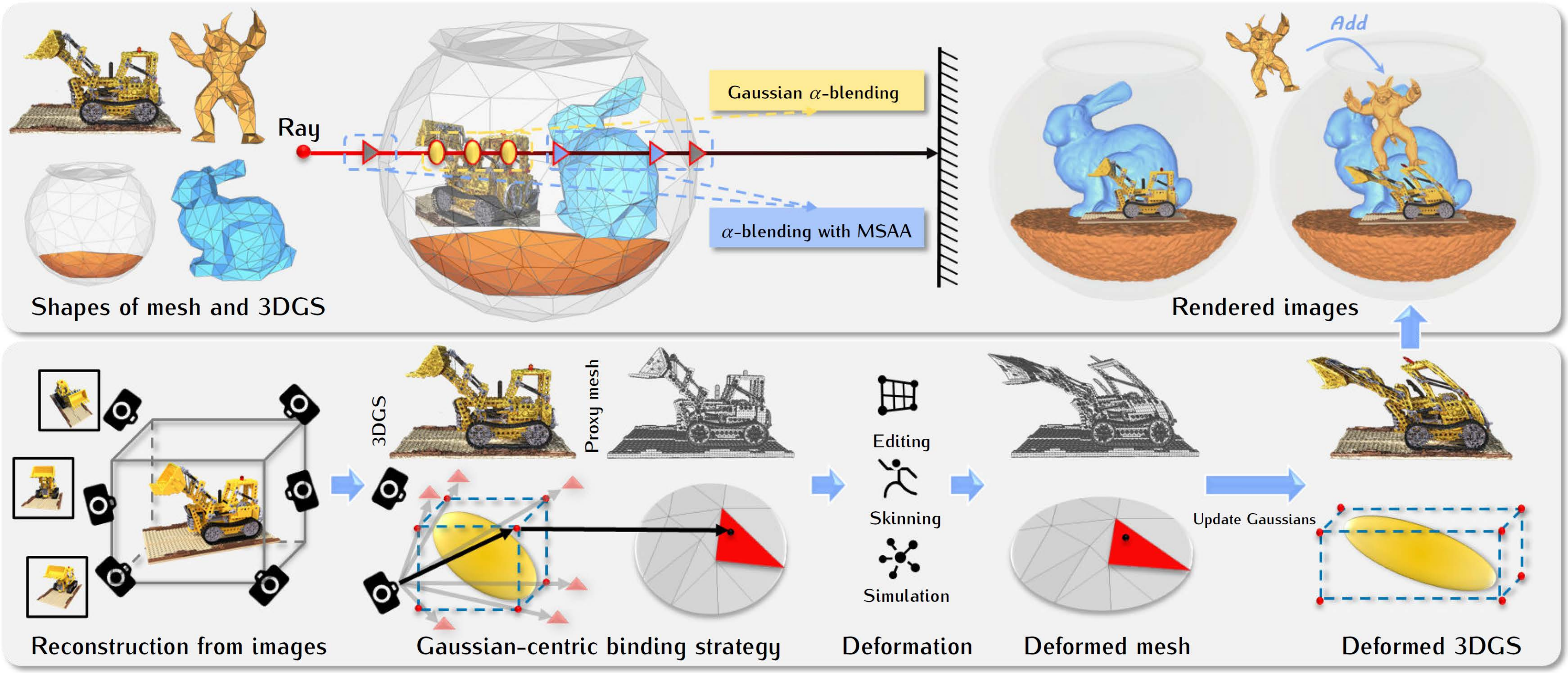}
    \end{minipage}
\caption{
Overview of UniMGS.
Given objects composed of 3DGS and mesh, the unified rendering pipeline bridges Gaussians and triangles by $\alpha$-blending in a single-pass manner, thus accurately computing color and handling occlusion.
To mitigate aliasing artifacts, we group depth-adjacent triangle fragments as a single entity, in which $\alpha$-blending with MSAA is performed.
The framework further allows mesh-based deformation to be seamlessly extended to 3DGS with a proxy mesh.
We associate a Gaussian with triangle faces by projecting the vertices of its BBX onto the mesh. During deformation, the motion of triangle faces is first transferred to the BBX and then propagated to the Gaussian.
}
\label{fig:sec3-framework}
\end{figure*}
\subsection{Representation for 3DGS Deformation}
3DGS~\cite{3dgs} is a discrete shape representation with each Gaussian continuously parameterized by the kernel function.
Though directly manipulating 3DGS is feasible~\cite{physgaussian,4dgs}, deformations often suffer from artifacts due to the lack of topological information.
Therefore, recent studies have proposed proxy-based representations that enable indirect deformation of 3DGS, such as sparse control points~\cite{scgs,MSGS,dreammesh4d}, deformation graph~\cite{arapgs}, cage~\cite{cageGS1,cageGS2}, and mesh~\cite{gaussianmesh,games}.
Among them, the proxy mesh performs better due to its complete topology and compatibility with most mesh-oriented manipulation methods, as confirmed by recent advanced research~\cite{gaussianmesh, manigs}.
Given a proxy mesh, SuGaR~\cite{sugar} and GaMeS~\cite{games} attach multiple Gaussians to each triangle face, then these Gaussians move with the triangle vertices weighted by barycentric coordinates.
Frosting~\cite{Frosting} forms a frosting layer around the mesh, where prismatic cells are created to enclose the Gaussians and control the Gaussians' motion during deformation.
Mani-GS~\cite{manigs} places Gaussians in the local coordinate system of mesh faces and drives the Gaussians through mesh editing.
GaussianMesh~\cite{gaussianmesh} assigns a single Gaussian to each triangle face, while they split together during training, and performs ACAP deformation~\cite{acap} on the mesh to edit Gaussians.
The above methods first initialize Gaussians on mesh faces through a mesh-centric binding scheme, then perform training, which makes the deformed Gaussians prone to visual artifacts due to their sensitivity to the topology quality of the proxy mesh.

Departing from previous studies, we propose a single-pass rasterization pipeline for 3DGS and mesh based on $\alpha$-blending, which facilitates transparency rendering and correct spatial occlusion relationship without dropping rendering rates.
In terms of proxy-based 3DGS deformation, we suggest a different insight by proposing a Gaussian-centric strategy that directly links trained 3DGS to the proxy mesh, resulting in minimal artifacts and robustness to mesh quality.
Collectively, these contributions promote significant progress in the visualization and manipulation of mesh and 3DGS within dynamic scenes.




\section{Methodology}

\subsection{Framework Overview}
Figure \ref{fig:sec3-framework} outlines our unified framework, which consists of two key modules: rasterization and deformation. 
We begin by reconstructing the 3DGS from multi-view images, while the corresponding proxy mesh can be obtained through various reconstruction algorithms~\cite{neus2,gof,poisson}.
Then, we apply the Gaussian-centric binding strategy to associate each Gaussian with relevant mesh faces. 
The deformation is initially performed on the proxy mesh and subsequently propagated to the 3DGS.
For rasterization of different objects represented by 3DGS or meshes, we integrate anti-aliased triangle $\alpha$-blending to the 3DGS rendering pipeline. 
This single-pass pipeline ensures correct handling of occlusion and color blending.
Next, we introduce the rasterization and deformation modules in detail.

\subsection{Rasterizing 3DGS and Mesh in a Single Pass}
\label{sec-4.2-rendering}
\begin{figure}[tb]
\centering
    \begin{subfigure}{0.21\textwidth}
        \centering
        \includegraphics[width=\textwidth]{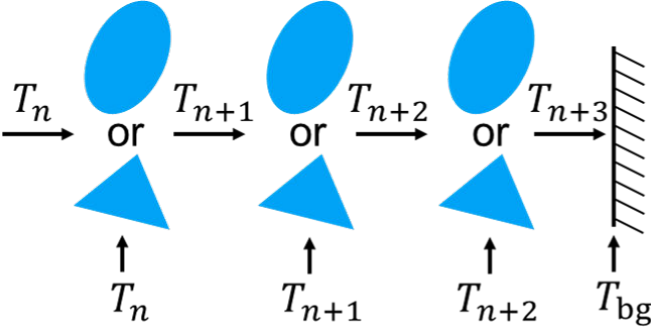}
        \caption{Direct $\alpha$-blending}
        \label{fig:sec3.2-msaa-a}
    \end{subfigure}%
 \hfill
    \begin{subfigure}{0.21\textwidth}
        \centering
        \includegraphics[width=\textwidth]{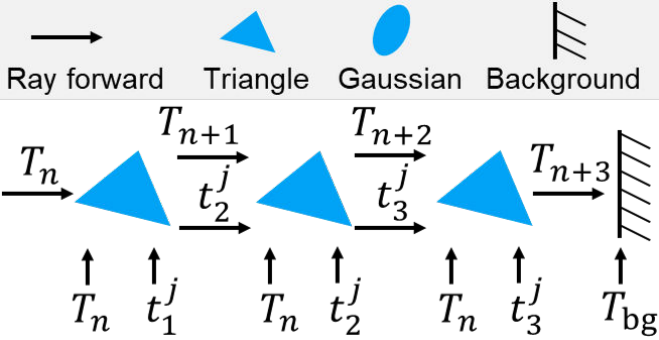}
        \caption{$\alpha$-blending with MSAA}
        \label{fig:sec3.2-msaa-c}
    \end{subfigure}%
 \hfill
    \begin{subfigure}{0.47\textwidth}
        \centering
        \includegraphics[width=\textwidth]{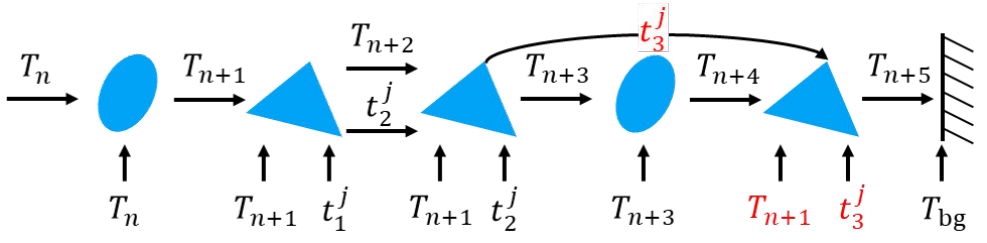}
        \caption{Color overflow}
        \label{fig:sec3.2-msaa-b}
    \end{subfigure}%
 \hfill
    \begin{subfigure}{0.47\textwidth}
        \centering
        \includegraphics[width=\textwidth]{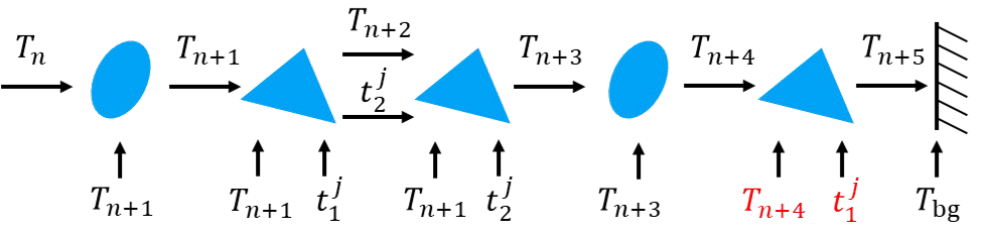}
        \caption{Our full method}
        \label{fig:sec3.2-msaa-d}
    \end{subfigure}%
        \caption{Illustration of single-pass rasterization with $\alpha$-blending.
        $\to$ means the update of transmittance while $\uparrow$ represents the transmittance used in $\alpha$-blending.
        (a): Directly blend triangles and Gaussians without anti-aliasing of triangles.
        (b): Consider all triangles overlapping the same pixel as a whole and perform MSAA in $\alpha$-blending.
        (c): Apply (b) directly to (a) causes color overflow.
        (d): Modify (c) by treating only depth-adjacent triangles as an entity, where the difference is highlighted in red.
        }
   	\label{fig:sec3.2-msaa}
\end{figure}
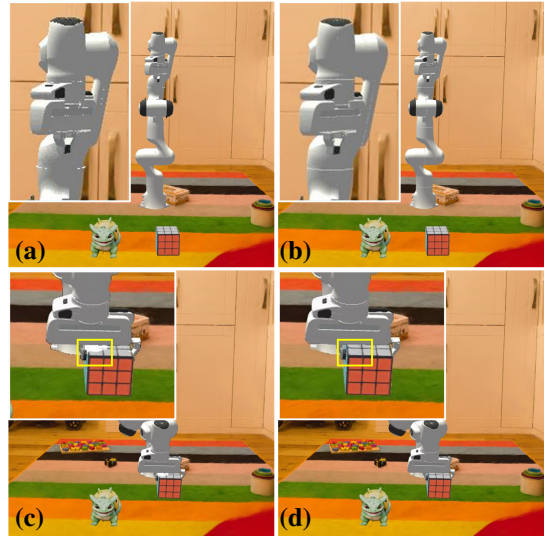
\begin{figure}[tb]
\centering
    \begin{minipage}[c]{0.4\textwidth}
    \centering
    \includegraphics[width=1\textwidth]{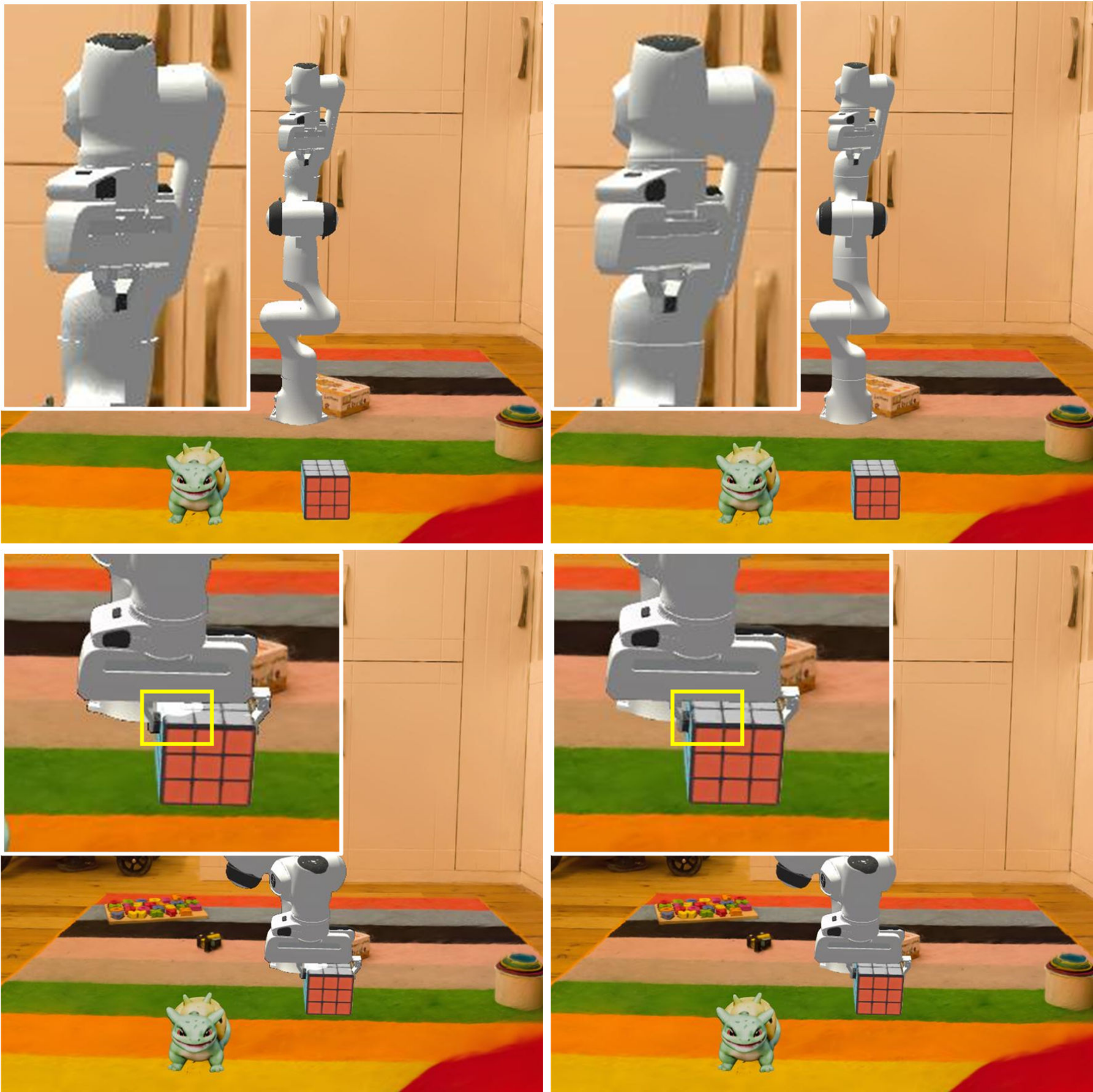}
    \end{minipage}
    \put(-200,6){\bf \textcolor{black}{(a)}}
    \put(-100,6){\bf \textcolor{black}{(b)}}
    \put(-200,-95){\bf \textcolor{black}{(c)}}
    \put(-100,-95){\bf \textcolor{black}{(d)}}
\caption{Visual improvements brought by our rasterization pipeline.
(a): Direct $\alpha$-blending; (b): Aliasing removed by combining MSAA with $\alpha$-blending; (c): Color overflow caused by incorrect transmittance; (d): Our full method without artifacts.
%
}
\label{fig:sec4.1-abl}
\end{figure}
To better clarify our contribution on rasterization, we provide additional diagrams and rendering results alongside the equations.
Figure \ref{fig:sec3.2-msaa} depicts how our unified rendering method is constructed, while Figure \ref{fig:sec4.1-abl} visualizes each part to justify the rationale behind each step of our method.

The 3DGS rasterizer employs $\alpha$-blending that the color $C_{\text{gs}}$ of a pixel is computed by blending $N$ depth-ordered fragments overlapping the pixel along a ray~\cite{3dgs}:
\begin{equation}
\label{eq:gs-alpha-blending}
    C_{\text{gs}} = \sum_{i=1}^N T_i\alpha_i c_i,
\end{equation}
with
\begin{equation}
\label{eq:gs-alpha-blending-transmittance}
    T_i = \prod_{j=1}^{i-1}(1-\alpha_j) \;\text{and}\; T_1=T_{\text{in}}.
\end{equation}
In Equation (\ref{eq:gs-alpha-blending}), $c_i$ and $\alpha_i$ are the RGB value and the opacity of the $i$-th fragment, $T_i$ is called transmittance and represents the visibility of the $i$-th fragment~\cite{renderSurvy}, and $T_{\text{in}}$ denotes the transmittance before the ray passes through these fragments, typically initialized to 1.0 if no fragments have been traversed.
For unified rasterization, a straightforward approach is to incorporate triangle fragments into the depth-sorting process, enabling seamless integration into the 3DGS rasterizer (see Figure \ref{fig:sec3.2-msaa-a} and \ref{fig:sec4.1-abl}a). 
However, while Gaussians alleviate aliasing through EWA filtering, it is also essential to equip triangle fragments with an anti-aliasing strategy.
Multi-Sample Anti-Aliasing (MSAA)~\cite{OpenGLbook} is a popular option~\cite{UnrealDoc} that scales an opaque triangle fragment’s color contribution by its coverage on sub-pixels:
\begin{equation}
\label{eq:msaa-opaque}
    C_{\text{tris}} = O_ic_i,
\end{equation}
with
\begin{equation}
    \label{eq:depth-test}
    O_i = \frac{1}{M}\sum_{j=1}^{M}\hat{o}_i^j\;\text{and}\;
    \hat{o}_{i}^{j}=\left\{
	\begin{aligned}
		o_i^j & , & o_{i-1} =0\\
		0 & , & o_{i-1} =1
	\end{aligned}\;,
	\right.
\end{equation}
where \smash{$o_i^j\in\{0,1\}$} represents whether the $i$-th fragment geometrically covers the $j$-th sub-pixel, \smash{$\hat{o}_i^j$} is the coverage after depth-testing, $O_i$ is the total coverage at pixel level, and M is the number of sub-pixels (we set M=4).
As mentioned in~\cite{renderSurvy}, a common approach combines MSAA with $\alpha$-blending by:
\begin{equation}
\label{eq:alpha-blending-msaa}
    C_{\text{tris}} = \sum_{i=1}^{N}T_iO_i\alpha_ic_i,
\end{equation}
with
\begin{equation}
\label{eq:alpha-blending-msaa-transmittance}
    O_i=\frac{1}{M}\sum_{j=1}^{M}o_i^j\; \text{and}\;T_i=\prod_{k=1}^{i-1}(1-O_k\alpha_k).
\end{equation}
In Equation (\ref{eq:alpha-blending-msaa-transmittance}), the visibility of a fragment depends on the transmission $T_i$ at the pixel level, but this should be performed at the sub-pixel level according to Equation (\ref{eq:depth-test}).
Since the detail of the transmittance calculation for each sub-pixel is not included in~\cite{renderSurvy}, we define \smash{$t_i^j$} within these triangle fragments according to Equation (\ref{eq:gs-alpha-blending-transmittance}):
\begin{equation}
\label{eq:t-check}
    t_i^j=\prod_{k=1}^{i-1}(1-o_k^j\alpha_k) \;\text{and}\; t_1^j=1.0.
\end{equation}
Thus, $O_i$ in Equation (\ref{eq:alpha-blending-msaa-transmittance}) is rewritten as:
\begin{equation}
\label{eq:alpha-blending-msaa-transmittance-new}
    O_i=\frac{1}{M}\sum_{j=1}^{M}o_i^jt_i^j.
\end{equation}
\smash{$O_i$} can be regarded as a continuous coverage influenced by the sub-pixel transmittance \smash{$t_i^j$} through weighted averaging, differing from the $M$ discrete values in Equation (\ref{eq:depth-test}).
Therefore, $T_i$ is no longer required, as transmittance is now handled at the sub-pixel level to accommodate the needs of MSAA.
As a result, we modify Equation (\ref{eq:alpha-blending-msaa}) to:
\begin{equation}
\label{eq:ours-triangle}
    C_{\text{tris}} = T_{\text{in}}\sum_{i=1}^{N}O_i\alpha_ic_i.
\end{equation}
Notably, though $T_i$ is dropped compared to Equation (\ref{eq:alpha-blending-msaa}), it is still updated by Equation (\ref{eq:alpha-blending-msaa-transmittance}) for the subsequent blending.
For example (see Figure \ref{fig:sec3.2-msaa-c} and \ref{fig:sec4.1-abl}b), given a background with \smash{$C_{\text{bg}}$} and \smash{$\alpha_{\text{bg}}$}, the final color of the pixel is:
\begin{equation}
\label{eq:bg-blending}
    C_{\text{pixel}} = C_{\text{tris}} + T_{\text{bg}}C_{\text{bg}},
\end{equation}
with 
\begin{equation}
\label{eq:bg-transmittance}
    T_{\text{bg}} = T_N(1-\alpha_{\text{bg}}).
\end{equation}
The main philosophy behind Equation (\ref{eq:ours-triangle}) is treating the $N$ triangles overlapping the same pixel as a single entity, so the updated transmittance \smash{$T_N$} at the pixel level contributes correctly to Equation (\ref{eq:bg-transmittance}).
However, this solution can not directly integrate with Gaussians.
When there are Gaussians between triangles, the transmittance used in $\alpha$-blending for triangles whose depth value is larger than Gaussians is too large (see the rightmost triangle in Figure \ref{fig:sec3.2-msaa-b}), as it does not consider the transmittance decay caused by the ray passing through Gaussians. This results in \textit{color overflow} and manifests as white artifacts (see Figure \ref{fig:sec4.1-abl}c).
To avoid \textit{color overflow}, we propose a simple yet effective method, where only the depth-adjacent triangles are treated as a single entity (see the two left triangles in Figure \ref{fig:sec3.2-msaa-d}), rather than all triangles overlapping the pixel.
During the unified rasterization of mesh and 3DGS, Equation (\ref{eq:gs-alpha-blending}) and Equation (\ref{eq:ours-triangle}) are selected for Gaussian and triangular fragments, respectively, which is in a single-pass anti-aliased manner (see Figure \ref{fig:sec4.1-abl}d).

\subsection{Deformation with Gaussian-Centric Binding}
\label{sec-binding}
\subsubsection{Binding Gaussians to Mesh.}
\label{sec3.3.1:binding}
We advocate a Gaussian-centric perspective and employ ray casting to associate Gaussians with the mesh faces, which is training-free and robust to mesh topology.
Since the cameras used in 3DGS training are inherently oriented toward the object, we cast rays from these cameras toward the center of a Gaussian.
If a ray hits a face, we record the face index, the barycentric coordinate of the intersection point, and the distance from the intersection point to the Gaussian center.
The Gaussian is then bound to the nearest candidate face.
However, this is not an ideal choice, as the size mismatch between Gaussians and faces can be significant, potentially leading to inconsistent deformation magnitudes.
Given that the cage or box deformation is robust~\cite{cageGS1}, we extend the Gaussian transformation to the average of the transformations of its BBX corners.
To this end, we propose an extended strategy where each camera casts 8 rays toward the corners of a Gaussian’s BBX. 
For each ray, we retain the face closest to the Gaussian. 
As a result, each Gaussian is ultimately bound to 8 faces.
This method is implemented using OptiX~\cite{optix}. Since ray casting is fully parallelized, the binding process completes within a few seconds.

\subsubsection{Deformation Transfer.}
\label{sec3.3.2:ACAP}
As we bind 3DGS to the proxy mesh, any deformation method for meshes can be extended to 3DGS, such as physical simulation and modifiers from the graphics engine.
Here, we briefly describe how deformation is transferred from the proxy mesh to 3DGS.
Consider a Gaussian kernel \smash{$G$} with its BBX \smash{$B$}, parameterized by a mean vector $\mu$ and a covariance matrix \smash{$\Sigma$}.
For notational clarity, the subscript \smash{$i=\{1,2,\cdots,8\}$} in any quantity \smash{$ X_i^j$} refers to the index of a BBX corner, while the superscript \smash{$j=\{1,2,3\}$} specifies the vertex indices of the associated triangle face.
Following GaussianMesh~\cite{gaussianmesh}, the ACAP~\cite{acap} is employed to calculate the transformation of a vertex during manipulation.
Specifically, the motion of a vertex consists of an offset \smash{$\Delta_i^j$}, a transformation matrix \smash{$D_i^j$} that can be decomposed into a rotation matrix \smash{$R_i^j$} and a shear matrix \smash{$S_i^j$} via polar decomposition.
Moreover, barycentric interpolation allows us to propagate the transformation from triangle vertices to any point within the triangle, which in our paper corresponds to the intersection point \smash{$P_i$} with barycentric coordinates \smash{$\{u,v,w\}$}.
The transformation of \smash{$P_i$} can be computed as follows:
\begin{equation}
    \begin{aligned}
                \Delta_i &= u\Delta_i^1 + v\Delta_i^2 + w\Delta_i^3\\
                &= u(V_i^{1\prime} - V_i^1) +  v(V_i^{2\prime} - V_i^2) +  w(V_i^{3\prime} - V_i^3), \\
        R_i &= u \log(R_i^1) + v \log(R_i^2) + w \log(R_i^3), \;\text{and}\\
        S_i &= u S_i^1 + v S_i^2 + w S_i^3,
    \end{aligned}
    \label{ACAP-bary}
\end{equation}
where \smash{$V_i^{j}$} and \smash{$V_i^{j\prime}$} represent the position of the \smash{$j$}-th vertex of a triangle before and after manipulation, respectively.
Since the transformations of the \smash{$i$}-th corner of \smash{$B$} keep the same with \smash{$P_i$}, they are subsequently transferred to $G$ by averaging operations:
\begin{equation}
\begin{aligned}
R^\prime &= \exp{(\frac{1}{8}\sum_{i=1}^8 R_i)},\; 
S^\prime = \frac{1}{8} \sum_{i=1}^8 S_i, \;
\Sigma^{\prime} = R^\prime S^\prime \Sigma (R^\prime S^\prime)^T,\\ 
\mu^{\prime} &= \mu + \frac{1}{8} \sum_{i=1}^8 \Delta _i,\;
\text{and}\;
G^{\prime}(x) = e^{ -\frac{1}{2}(x-\mu^{\prime})^T(\Sigma^{\prime})^{-1}(x-\mu^{\prime})},
\end{aligned}
\label{ACAP-avg}
\end{equation}
where \smash{$G^\prime(x)$} is the updated Gaussian kernel. 
For more details, please see the references~\cite{gaussianmesh,acap}.

\begin{figure*}[t]
\centering
    \begin{subfigure}{0.45\textwidth}
        \centering
        \includegraphics[width=1\textwidth]{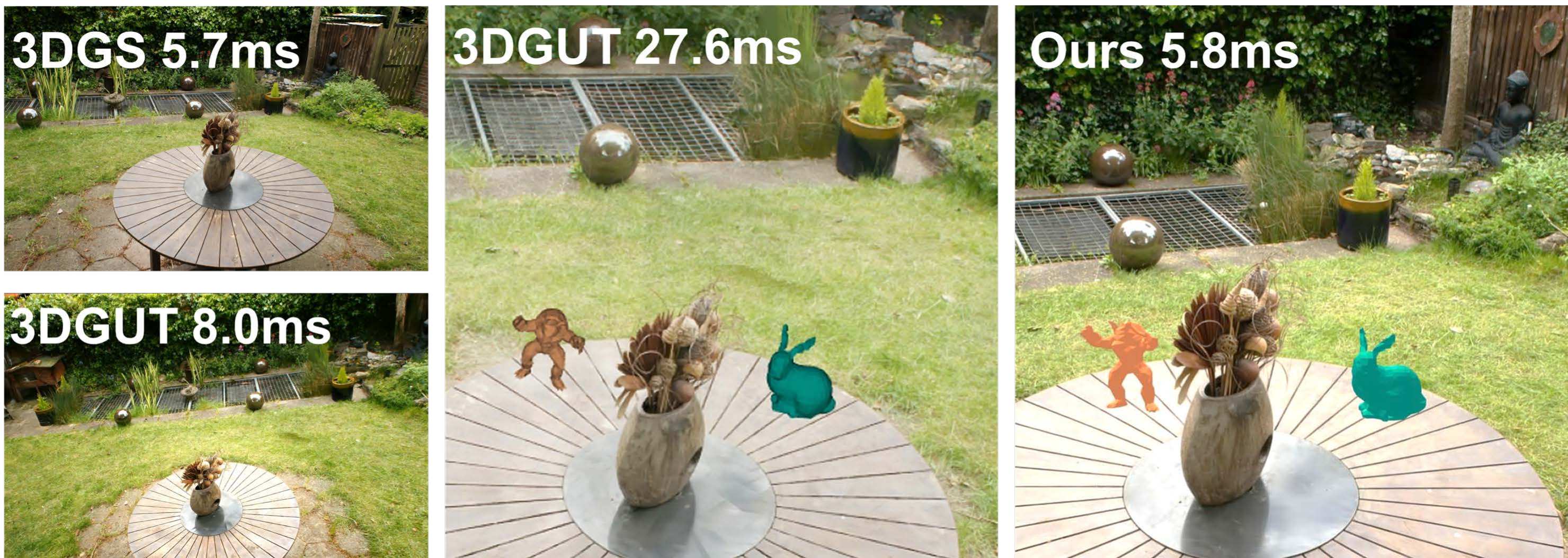}
        \caption{Comparing with ray-based methods}
        \label{fig:sec4.1-iccv}
    \end{subfigure}%
 \hfill
    \begin{subfigure}{0.50\textwidth}
        \centering
        \includegraphics[width=1\textwidth]{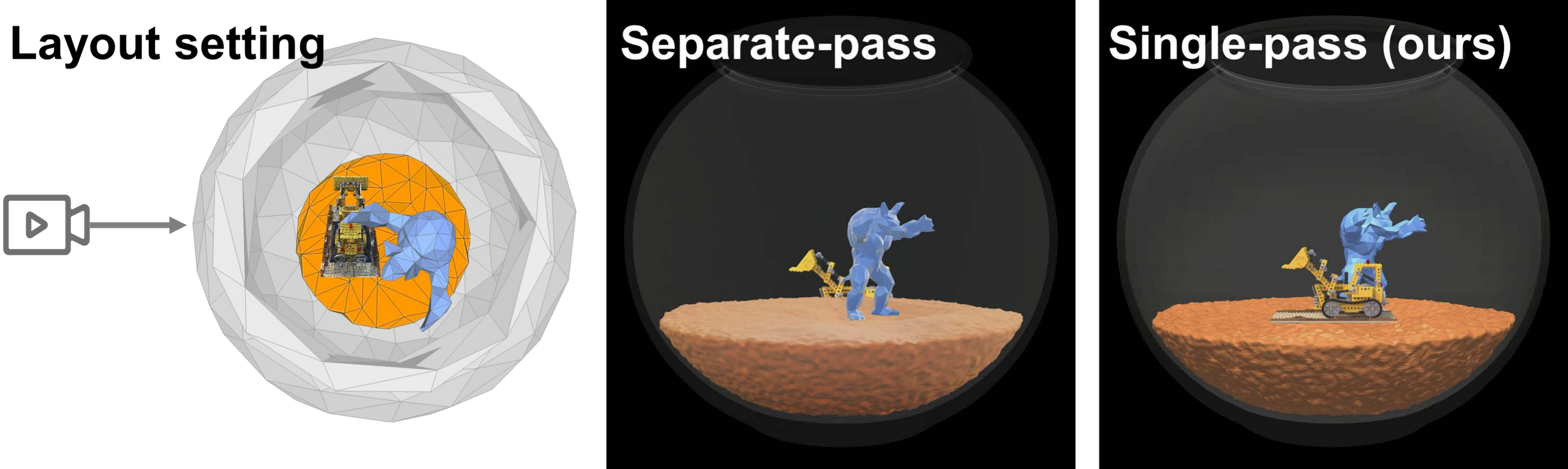}
        \caption{Comparing with separate-pass scheme}
        \label{fig:sec4.1-separate}
    \end{subfigure}%
\caption{
Evaluation against existing hybrid rendering studies. 
(a): Compared to ray-based methods, our method shows promising runtime performance while maintaining comparable visual effects.
(b): Compared to separate-pass rasterization work, our method guarantees the correct spatial relationship because the colors of Gaussians and triangles are blended in a single pass.
}
\label{fig:sec4.1-comparision}
\end{figure*}

\section{Experiments}

We comprehensively evaluate the capabilities of \textit{UniMGS} through quantitative and qualitative experiments, including the performance of our unified rasterization, the improvement in deformation quality brought by the Gaussian-centric binding strategy, and two representative applications based on our unified framework.
For our method, all Gaussian objects are reconstructed using 3DGS~\cite{3dgs}, while for others, we follow their official instructions to train them until convergence.
Evaluations are based on the public dataset of NeRF-Synthetic~\cite{nerf}, MipNeRF360~\cite{barron2022mip}, Hybrid-IBR~\cite{hybridIBR}, and our self-collected data from Fab~\cite{fab3d}.
Due to space limitations, we only show a subset of the results. 
Additional examples are provided in the supplementary material.

{\fontsize{9}{11}\selectfont
\begin{table}[tb]
\centering
\begin{tabular}{lcccc}
\toprule
Method & PSNR$\uparrow$ & SSIM$\uparrow$ & LPIPS$\downarrow$ & \textit{Count} \\ 
\midrule
\textit{UniMGS}         & \textbf{28.43} & \textbf{0.946} & \textbf{0.048} & 275K\\
\textit{UniMGS* }       & \underline{28.23} & \underline{0.940} & \underline{0.051} & 275K\\
GaussianMesh   & 27.21 & 0.925 & 0.079 & 989K\\
Frosting       & 25.42 & 0.912 & 0.059 & 2.16M\\
Mani-GS        & 23.76 & 0.842 & 0.123 & 2.47M\\
\bottomrule
\end{tabular}
\caption{Quantitative comparison of novel view synthesis on NeRF-Synthetic dataset.}
\label{tab:psnr}
\end{table}
}

\subsection{Unified Rasterization Ability}

\subsubsection{Baselines.}
Existing studies~\cite{xie2025vid2sim,mega,bridging} on coupled rasterization of Gaussians and meshes all follow a separate-pass paradigm, but none of them are available.
Therefore, we have to reproduce the separate-pass rendering scheme for an intuitive evaluation.
In addition, 3DGUT~\cite{3dgut} employs ray-based rendering for 3DGS and meshes.
Implementation details are provided in the supplementary material.
\subsubsection{Results.} 
In Figure \ref{fig:sec4.1-iccv}, we designed a benchmark scene composed of Gaussians (\textit{garden} from MipNeRF360~\cite{barron2022mip}) and low-poly meshes (shapes in orange and green), and recorded the average per-frame rendering time.
With Gaussians only, 3DGUT takes 1.4× longer to render than 3DGS, consistent with its original findings.
When meshes are added, our rendering speed remains stable, while 3DGUT slows down by 3.4×.
This performance drop is attributed to 3DGUT’s reliance on additional ray-tracing passes for meshes, while our approach employs a unified rasterization pipeline that handles both meshes and Gaussians in a single pass.
Figure \ref{fig:sec4.1-separate} depicts a scenario in which a \textit{lego} represented by Gaussians is placed alongside mesh objects, enclosed by a transparent bowl.
However, the separate-pass approach fails to handle occlusion in this case.
In the separate-pass scheme, the \textit{lego} and mesh images are composited pixel-wise based on depth. However, due to the nested spatial relationship, the \textit{lego} is impossible to appear inside the bowl.
The key reason is that correct visibility can only be ensured within the same representation, while depth maps alone are insufficient to guarantee the correctness of subsequent color composition.
In contrast, our single-pass method achieves true unified rendering by computing color at the fragment level.

\subsection{Deformation Performance}
This experiment evaluates the deformation performance of mesh-centric and our Gaussian-centric binding strategies across different animation algorithms and proxy mesh sources.
Visual artifacts serve as a primary indicator.
Several advanced deformation representations have been proposed by GaussianMesh~\cite{gaussianmesh}, Mani-GS~\cite{manigs}, and Frosting~\cite{Frosting}, which are mesh-centric methods.
For our method, we refer to the direct bind of a Gaussian to a single face as \textit{UniMGS*}, and the BBX-based association to multiple faces as \textit{UniMGS}.

\subsubsection{Quantitative Comparison on NeRF-Synthetic Dataset.}
%
In the NeRF-Synthetic dataset~\cite{nerf}, the artist-created Ground Truth (GT) meshes are employed as proxies, whose triangle face count ranges from 60K to 1,200K.
Following PhysGaussian~\cite{physgaussian}, we perform deformation with the Simple Deform Modifier in Blender~\cite{blender} and evaluate the results using PSNR, SSIM, and LPIPS.
These metrics are computed by comparing the novel view renderings of the deformed mesh and 3DGS.
In addition, the average Gaussian count is reported as \textit{Count}.

\begin{figure*}[t]
\centering
    \begin{minipage}[c]{1\textwidth}
    \centering
    \includegraphics[width=0.95\textwidth]{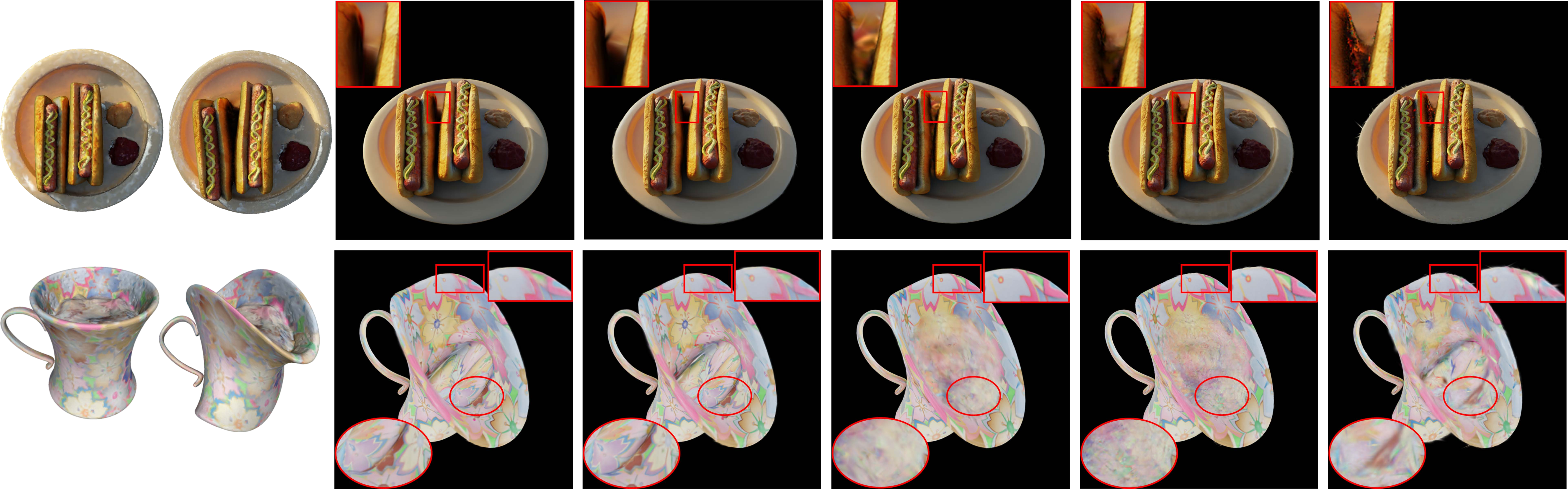}
    \put(-460,152){\bf  Proxy mesh}
    \put(-356,152){\bf UniMGS}
    \put(-285,152){\bf UniMGS*}
    \put(-220,152){\bf GaussianMesh}
    \put(-130,152){\bf Frosting}
    \put(-55,152){\bf Mani-GS}
    \put(-490,100){\bf \rotatebox{90}{GOF}}
    \put(-490,30){\bf \rotatebox{90}{NeuS2}}
    \put(-440,80){\small \bf \textit{hotdog}}
    \put(-440,5){\small \bf \textit{cup}}
    \put(-327,140){\small \bf \textcolor{yellow}{166K}}
    \put(-250,140){\small \bf \textcolor{yellow}{166K}}
    \put(-173,140){\small \bf \textcolor{yellow}{332K}}
    \put(-96,140){\small \bf \textcolor{yellow}{2.0M}}
    \put(-22,140){\small \bf \textcolor{yellow}{150K}}
    \put(-327,3){\small \bf \textcolor{yellow}{281K}}
    \put(-250,3){\small \bf \textcolor{yellow}{281K}}
    \put(-173,3){\small \bf \textcolor{yellow}{247K}}
    \put(-96,3){\small \bf \textcolor{yellow}{2.0M}}
    \put(-22,3){\small \bf \textcolor{yellow}{150K}}
    \end{minipage}
\caption{Visual comparison of deformation.
We visualize textured proxy meshes before and after deformation, obtained from two advanced reconstruction methods (NeuS2 and GOF), where the number of Gaussians is marked in yellow.
All proxy meshes are simplified to 50K faces.
The baselines (GaussianMesh, Frosting, and Mani-GS) suffer from severe artifacts under large deformations, primarily due to their face-centric binding strategies, which are inherently sensitive to the quality of the proxy mesh topology.
}
\label{fig:sec4.2-deformation}
\end{figure*}
\begin{figure*}[tb]
\centering
    \begin{minipage}[c]{0.98\textwidth}
    \centering
    \includegraphics[width=1\textwidth]{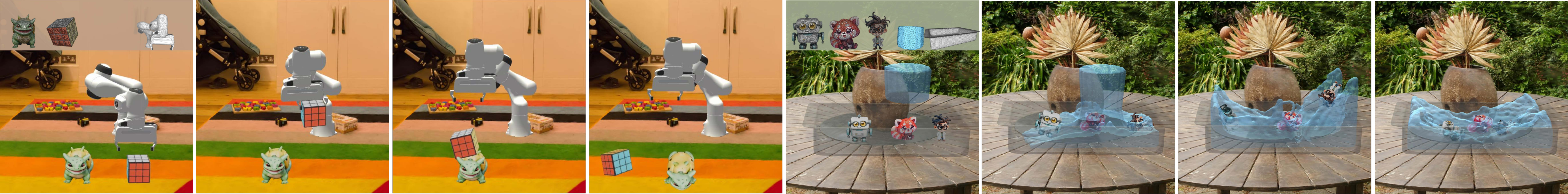}
    \put(-490,57){\tiny \textcolor{white}{3DGS}}
    \put(-452,57){\tiny \textcolor{white}{Mesh}}
    \put(-237,57){\tiny \textcolor{white}{3DGS}}
    \put(-207,57){\tiny \textcolor{white}{Mesh}}
    \end{minipage}
\caption{
Applications in embodied AI and fluid simulation.
These two scenarios involve complex occlusions, rich interactions, and diverse appearances (including both transparent and opaque objects), which are all faithfully presented.
The backgrounds are reconstructed by 3DGS~\cite{3dgs}.
%
}
\label{fig:sec4.3-demo}
\end{figure*}
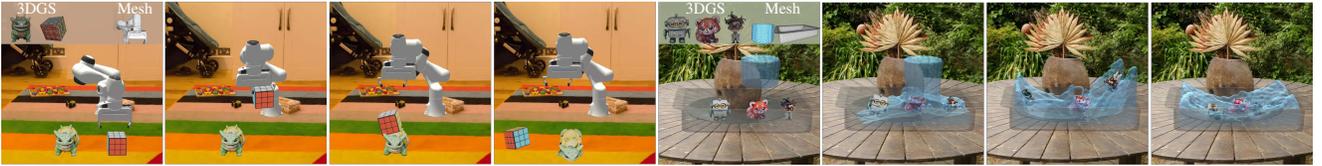
As reported in Table \ref{tab:psnr}, our method outperforms others in all quantitative metrics, where the corresponding visualizations are provided in the supplementary material.
In GaussianMesh, faces and Gaussians could be subdivided together during training; Frosting introduces an additional frosting layer composed of Gaussians; and Mani-GS assigns multiple Gaussians to each face.
These mesh-centric strategies must bind Gaussians to the mesh before training, guided by geometric constraints from the proxy mesh. While beneficial for static 3DGS optimization, such constraints may impair deformation quality and produce artifacts.
Instead, our Gaussian-centric strategy directly links the trained 3DGS to the mesh, thus keeping the original 3DGS data unchanged, while preserving promising results.

\subsubsection{Robustness to Mesh Quality.}
Since high-quality meshes are not always available, the proxy meshes in this experiment are obtained from NeuS2~\cite{neus2} (recommended by GaussianMesh and Mani-GS) and GOF~\cite{gof}.
Considering computational cost, we follow GaussianMesh~\cite{gaussianmesh} and simplify all proxy meshes to 50K faces using the decimation method~\cite{qem,meshlab}.
Physical simulation~\cite{macklin201xpbd} is imposed on \textit{hotdog}, whereas Blender modifiers are employed for \textit{cup}.
Figure \ref{fig:sec4.2-deformation} illustrates the proxy meshes before and after deformation, along with the resulting deformed 3DGS.
GOF is an advanced method for extracting textured meshes from 3DGS, while NeuS2 relies on implicit neural fields and consistently generates watertight meshes.
However, NeuS2 exhibits a failure case that erroneously closes the opening of the \textit{cup}.
Under such an unfavorable condition, existing mesh-centric methods perform poorly in Gaussian optimization, as Gaussians are constrained near the proxy mesh surface.
In contrast, our Gaussian-centric approach spatially links Gaussians to the mesh without imposing a causal relationship, thus offering greater tolerance even when the proxy mesh is of poor quality or structurally flawed.
Disregarding mesh quality, GaussianMesh also adopts ACAP for deformation transfer, but suffers from distortion and blurring, further highlighting the advantage of our Gaussian-centric binding strategy.
Compared to \textit{UniMGS*}, the introduction of BBX in \textit{UniMGS} improves visual performance.

\subsection{Applications Based on UniMGS}
UniMGS unifies mesh and 3DGS in both rasterization and deformation, laying the foundation for downstream tasks involving hybrid representations.
Figure~\ref{fig:sec4.3-demo} shows two representative cases to demonstrate the application potential of UniMGS in embodied AI and fluid simulation.
Implementation details are provided in the supplementary material.

\section{Conclusion}
In this paper, we first address the underexplored problem of unified rasterization of 3DGS and mesh by blending the colors of both triangle and Gaussian fragments in a single-pass manner, ensuring accurate color computation and occlusion handling.
Second, we propose a novel Gaussian-centric binding strategy to enhance the visual performance of 3DGS driven by proxy meshes.
Finally, comprehensive experiments validate the superiority of \textit{UniMGS} in rendering and manipulation, while the application cases further emphasize the practical significance.

\bibliography{aaai2026}

@Inbook{CGBOOK,
  author    = {Hughes, John F. and van Dam, Andries and McGuire, Morgan and Sklar, David F. and Foley, James D. and Feiner, Steven K. and Akeley, Kurt},
  title     = {{Computer Graphics Principles and Practice (Inbook-text-in-chap)}},
  chapter   = {{Ray Casting and Raterization}},
  year      = {2013},
  address   = {Upper Saddle River, NJ, USA},
  publisher = {Addison-Wesley Professional},
  type      = {Chapter:},
  pages     = {387--393},
  numpages  = {7},
  isbn      = {978-0-321-39952-6},
}

@Inbook{OpenGLbook,
  author    = {Joey de Vries},
  title     = {{Learn OpenGL: Learn Modern OpenGL Graphics Programming in a Step-by-Step Fashion (Inbook-text-in-chap)}},
  chapter   = {{Part IV - Advanced OpenGL Anti Aliasing}},
  year      = {2020},
  publisher = {Kendall \& Welling},
  type      = {Chapter:},
  pages     = {264--271},
  numpages  = {8},
  isbn      = {978-90-90-33256-7},
}

@inproceedings{nerf,
  title      = {{NeRF: Representing Scenes as Neural Radiance Fields for View Synthesis}},
  author     = {Mildenhall, Ben and Srinivasan, Pratul P and Tancik, Matthew and Barron, Jonathan T and Ramamoorthi, Ravi and Ng, Ren},
  booktitle  = {{Proceedings of the European Conference on Computer Vision (ECCV)}},
  pages      = {405--421},
  year       = {2020},
  organization = {Springer}
}

@article{3dgs,
  title   = {{3D Gaussian Splatting for Real-Time Radiance Field Rendering}},
  author  = {Kerbl, Bernhard and Kopanas, Georgios and Leimk{\"u}hler, Thomas and Drettakis, George},
  journal = {ACM Trans. Graph.},
  volume  = {42},
  number  = {4},
  pages   = {139--1},
  year    = {2023}
}

@inproceedings{nerfmesh-iccv,
  title     = {{Dynamic Mesh-Aware Radiance Fields}},
  author    = {Qiao, Yi-Ling and Gao, Alexander and Xu, Yiran and Feng, Yue and Huang, Jia-Bin and Lin, Ming C},
  booktitle = {{Proceedings of the IEEE/CVF International Conference on Computer Vision (ICCV)}},
  pages     = {385--396},
  year      = {2023}
}

@article{nerfmesh-tvcg,
  title     = {{A Real-time Method for Inserting Virtual Objects into Neural Radiance Fields}},
  author    = {Ye, Keyang and Wu, Hongzhi and Tong, Xin and Zhou, Kun},
  journal   = {IEEE Transactions on Visualization and Computer Graphics},
  year      = {2024},
  publisher = {IEEE}
}

@article{gaussianmesh,
  title     = {{Real-time Large-Scale Deformation of Gaussian Splatting}},
  author    = {Gao, Lin and Yang, Jie and Zhang, Bo-Tao and Sun, Jia-Mu and Yuan, Yu-Jie and Fu, Hongbo and Lai, Yu-Kun},
  journal   = {ACM Trans. Graph.},
  volume    = {43},
  number    = {6},
  pages     = {1--17},
  year      = {2024},
  publisher = {ACM New York, NY, USA}
}

@inproceedings{4dgs,
  title     = {{4D Gaussian Splatting for Real-Time Dynamic Scene Rendering}},
  author    = {Wu, Guanjun and Yi, Taoran and Fang, Jiemin and Xie, Lingxi and Zhang, Xiaopeng and Wei, Wei and Liu, Wenyu and Tian, Qi and Wang, Xinggang},
  booktitle = {{Proceedings of the IEEE/CVF Conference on Computer Vision and Pattern Recognition (CVPR)}},
  pages     = {20310--20320},
  year      = {2024}
}

@inproceedings{physgaussian,
  title     = {{PhysGaussian: Physics-Integrated 3D Gaussians for Generative Dynamics}},
  author    = {Xie, Tianyi and Zong, Zeshun and Qiu, Yuxing and Li, Xuan and Feng, Yutao and Yang, Yin and Jiang, Chenfanfu},
  booktitle = {{Proceedings of the IEEE/CVF Conference on Computer Vision and Pattern Recognition (CVPR)}},
  pages     = {4389--4398},
  year      = {2024}
}

@inproceedings{scgs,
  title     = {{SC-GS: Sparse-Controlled Gaussian Splatting for Editable Dynamic Scenes}},
  author    = {Huang, Yi-Hua and Sun, Yang-Tian and Yang, Ziyi and Lyu, Xiaoyang and Cao, Yan-Pei and Qi, Xiaojuan},
  booktitle = {{Proceedings of the IEEE/CVF Conference on Computer Vision and Pattern Recognition (CVPR)}},
  pages     = {4220--4330},
  year      = {2024}
}

@article{dreammesh4d,
  title   = {{DreamMesh4D: Video-to-4D Generation with Sparse-Controlled Gaussian-Mesh Hybrid Representation}},
  author  = {Li, Zhiqi and Chen, Yiming and Liu, Peidong},
  journal = {Advances in Neural Information Processing Systems},
  volume  = {37},
  pages   = {21377--21400},
  year    = {2024}
}

@inproceedings{sugar,
  title     = {{SuGaR: Surface-Aligned Gaussian Splatting for Efficient 3D Mesh Reconstruction and High-Quality Mesh Rendering}},
  author    = {Gu{\'e}don, Antoine and Lepetit, Vincent},
  booktitle = {{Proceedings of the IEEE/CVF Conference on Computer Vision and Pattern Recognition (CVPR)}},
  pages     = {5354--5363},
  year      = {2024}
}

@article{games,
  title   = {{GAMES: Mesh-Based Adapting and Modification of Gaussian Splatting}},
  author  = {Waczy{\'n}ska, Joanna and Borycki, Piotr and Tadeja, S{\l}awomir and Tabor, Jacek and Spurek, Przemys{\l}aw},
  journal = {arXiv preprint arXiv:2402.01459},
  year    = {2024}
}

@article{manigs,
  title   = {{Mani-GS: Gaussian Splatting Manipulation with Triangular Mesh}},
  author  = {Gao, Xiangjun and Li, Xiaoyu and Zhuang, Yiyu and Zhang, Qi and Hu, Wenbo and Zhang, Chaopeng and Yao, Yao and Shan, Ying and Quan, Long},
  journal = {arXiv preprint arXiv:2405.17811},
  year    = {2024}
}

@inproceedings{Frosting,
  title        = {{Gaussian Frosting: Editable Complex Radiance Fields with Real-Time Rendering}},
  author       = {Gu{\'e}don, Antoine and Lepetit, Vincent},
  booktitle    = {{Proceedings of the European Conference on Computer Vision (ECCV)}},
  pages        = {413--430},
  year         = {2024},
  organization = {Springer}
}

@article{bridging,
  title   = {{Bridging 3D Gaussian and Mesh for Freeview Video Rendering}},
  author  = {Xiao, Yuting and Wang, Xuan and Li, Jiafei and Cai, Hongrui and Fan, Yanbo and Xue, Nan and Yang, Minghui and Shen, Yujun and Gao, Shenghua},
  journal = {arXiv preprint arXiv:2403.11453},
  year    = {2024}
}

@inproceedings{ewa2,
  title        = {{EWA Volume Splatting}},
  author       = {Zwicker, Matthias and Pfister, Hanspeter and Van Baar, Jeroen and Gross, Markus},
  booktitle    = {{Proceedings Visualization, 2001. VIS'01.}},
  pages        = {29--538},
  year         = {2001},
  organization = {IEEE}
}

@inproceedings{poisson,
  title     = {{Poisson Surface Reconstruction}},
  author    = {Kazhdan, Michael and Bolitho, Matthew and Hoppe, Hugues},
  booktitle = {{Proceedings of the Fourth Eurographics Symposium on Geometry Processing}},
  volume    = {7},
  year      = {2006}
}

@inproceedings{neus2,
  title     = {{Neus2: Fast Learning of Neural Implicit Surfaces for Multi-View Reconstruction}},
  author    = {Wang, Yiming and Han, Qin and Habermann, Marc and Daniilidis, Kostas and Theobalt, Christian and Liu, Lingjie},
  booktitle = {{Proceedings of the IEEE/CVF International Conference on Computer Vision (ICCV)}},
  pages     = {3295--3306},
  year      = {2023}
}

@article{gof,
  title     = {{Gaussian Opacity Fields: Efficient Adaptive Surface Reconstruction in Unbounded Scenes}},
  author    = {Yu, Zehao and Sattler, Torsten and Geiger, Andreas},
  journal   = {ACM Trans. Graph.},
  volume    = {43},
  number    = {6},
  pages     = {1--13},
  year      = {2024},
  publisher = {ACM New York, NY, USA}
}

@article{arapgs,
  title     = {{As-Rigid-As-Possible Deformation of Gaussian Radiance Fields}},
  author    = {Tong, Xinhao and Shao, Tianjia and Weng, Yanlin and Yang, Yin and Zhou, Kun},
  journal   = {IEEE Transactions on Visualization and Computer Graphics},
  year      = {2025},
  publisher = {IEEE}
}

@inproceedings{MSGS,
  title        = {{Reconstruction and Simulation of Elastic Objects with Spring-Mass 3D Gaussians}},
  author       = {Zhong, Licheng and Yu, Hong-Xing and Wu, Jiajun and Li, Yunzhu},
  booktitle    = {{Proceedings of the European Conference on Computer Vision (ECCV)}},
  pages        = {407--423},
  year         = {2024},
  organization = {Springer}
}

@article{cageGS1,
  title   = {{CAGE-GS: High-Fidelity Cage Based 3D Gaussian Splatting Deformation}},
  author  = {Tong, Yifei and Tian, Runze and Han, Xiao and Liu, Dingyao and Yu, Fenggen and Zhang, Yan},
  journal = {arXiv preprint arXiv:2504.12800},
  year    = {2025}
}

@article{cageGS2,
  title   = {{GSDeformer: Direct Cage-Based Deformation for 3D Gaussian Splatting}},
  author  = {Huang, Jiajun and Yu, Hongchuan},
  journal = {arXiv preprint arXiv:2405.15491},
  year    = {2024}
}

@article{xie2025vid2sim,
  title   = {{Vid2Sim: Realistic and Interactive Simulation from Video for Urban Navigation}},
  author  = {Xie, Ziyang and Liu, Zhizheng and Peng, Zhenghao and Wu, Wayne and Zhou, Bolei},
  journal = {arXiv preprint arXiv:2501.06693},
  year    = {2025}
}

@inproceedings{xia2024video2game,
  title     = {{Video2Game: Real-Time Interactive Realistic and Browser-Compatible Environment from a Single Video}},
  author    = {Xia, Hongchi and Lin, Zhi-Hao and Ma, Wei-Chiu and Wang, Shenlong},
  booktitle = {{Proceedings of the IEEE/CVF Conference on Computer Vision and Pattern Recognition (CVPR)}},
  pages     = {4578--4588},
  year      = {2024}
}

@inproceedings{macklin201xpbd,
  title     = {{XPBD: Position-Based Simulation of Compliant Constrained Dynamics}},
  author    = {Macklin, Miles and M{\"u}ller, Matthias and Chentanez, Nuttapong},
  booktitle = {{Proceedings of the 9th International Conference on Motion in Games}},
  pages     = {49--54},
  year      = {2016}
}

@inproceedings{renderSurvy,
  title        = {{Transparency and Anti-Aliasing Techniques for Real-Time Rendering}},
  author       = {Maule, Marilena and Comba, Joao LD and Torchelsen, Rafael and Bastos, Rui},
  booktitle    = {{2012 25th SIBGRAPI Conference on Graphics, Patterns and Images Tutorials}},
  pages        = {50--59},
  year         = {2012},
  organization = {IEEE}
}

@inproceedings{guo2023vmesh,
  title     = {{VMesh: Hybrid Volume-Mesh Representation for Efficient View Synthesis}},
  author    = {Guo, Yuan-Chen and Cao, Yan-Pei and Wang, Chen and He, Yu and Shan, Ying and Zhang, Song-Hai},
  booktitle = {{SIGGRAPH Asia 2023 Conference Papers}},
  pages     = {1--11},
  year      = {2023}
}

@misc{blender,
  author       = "{Blender Foundation}",
  title        = {{Blender 4.4 Manual}},
  year         = {2025},
  howpublished = "\url{https://www.blender.org/}",
  note         = "[Online; accessed 25-April-2025]"
}

@misc{fab3d,
  author       = "{Epic Games, Inc.}",
  title        = {{Fab}},
  year         = {2025},
  howpublished = "\url{https://www.fab.com/}",
  note         = "[Online; accessed 18-May-2025]"
}

@inproceedings{qem,
  title     = {{Surface Simplification Using Quadric Error Metrics}},
  author    = {Garland, Michael and Heckbert, Paul S},
  booktitle = {{Proceedings of the 24th Annual Conference on Computer Graphics and Interactive Techniques}},
  pages     = {209--216},
  year      = {1997}
}

@inproceedings{meshlab,
  booktitle = {Eurographics Italian Chapter Conference},
  editor    = {Vittorio Scarano and Rosario De Chiara and Ugo Erra},
  title     = {{MeshLab: an Open-Source Mesh Processing Tool}},
  author    = {Cignoni, Paolo and Callieri, Marco and Corsini, Massimiliano and Dellepiane, Matteo and Ganovelli, Fabio and Ranzuglia, Guido},
  year      = {2008},
  publisher = {The Eurographics Association},
  ISBN      = {978-3-905673-68-5},
  DOI       = {10.2312/LocalChapterEvents/ItalChap/ItalianChapConf2008/129-136}
}

@article{optix,
  author    = {Parker, Steven G. and Bigler, James and Dietrich, Andreas and Friedrich, Heiko and Hoberock, Jared and Luebke, David and McAllister, David and McGuire, Morgan and Morley, Keith and Robison, Austin and Stich, Martin},
  title     = {{OptiX: A General Purpose Ray Tracing Engine}},
  year      = {2010},
  issue_date = {July 2010},
  publisher = {Association for Computing Machinery},
  address   = {New York, NY, USA},
  volume    = {29},
  number    = {4},
  issn      = {0730-0301},
  url       = {https://doi.org/10.1145/1778765.1778803},
  doi       = {10.1145/1778765.1778803},
  journal   = {ACM Trans. Graph.},
  month     = Jul,
  articleno = {66},
  numpages  = {13}
}

@ARTICLE{acap,
  author    = {Gao, Lin and Lai, Yu-Kun and Yang, Jie and Zhang, Ling-Xiao and Xia, Shihong and Kobbelt, Leif},
  journal   = {IEEE Transactions on Visualization and Computer Graphics},
  title     = {{Sparse Data Driven Mesh Deformation}},
  year      = {2021},
  volume    = {27},
  number    = {3},
  pages     = {2085-2100},
  doi       = {10.1109/TVCG.2019.2941200}
}

@article{mega,
  title   = {{Mega: Hybrid Mesh-Gaussian Head Avatar for High-Fidelity Rendering and Head Editing}},
  author  = {Wang, Cong and Kang, Di and Sun, He-Yi and Qian, Shen-Han and Wang, Zi-Xuan and Bao, Linchao and Zhang, Song-Hai},
  journal = {arXiv preprint arXiv:2404.19026},
  year    = {2024}
}

@article{hybridIBR,
  title   = {{Hybrid Image-Based Rendering for Free-View Synthesis}},
  author  = {Prakash, Siddhant and Leimk{\"u}hler, Thomas and Rodriguez, Simon and Drettakis, George},
  journal = {Proceedings of the ACM on Computer Graphics and Interactive Techniques},
  volume  = {4},
  number  = {1},
  pages   = {1--20},
  year    = {2021}
}

@misc{UnrealDoc,
  author       = "{Epic Games, Inc.}",
  title        = {{Unreal Engine 5.6 Documentation: Anti-Aliasing and Upscaling}},
  year         = {2025},
  howpublished = "\url{https://dev.epicgames.com}",
  note         = "[Online; accessed 20-June-2025]"
}

@article{3dgrt,
  author  = {Nicolas Moenne-Loccoz and Ashkan Mirzaei and Or Perel and Riccardo de Lutio and Janick Martinez Esturo and Gavriel State and Sanja Fidler and Nicholas Sharp and Zan Gojcic},
  title   = {{3D Gaussian Ray Tracing: Fast Tracing of Particle Scenes}},
  journal = {ACM Transactions on Graphics (SIGGRAPH Asia)},
  year    = {2024},
}

@article{3dgut,
  title   = {{3DGUT: Enabling Distorted Cameras and Secondary Rays in Gaussian Splatting}},
  author  = {Wu, Qi and Martinez Esturo, Janick and Mirzaei, Ashkan and Moenne-Loccoz, Nicolas and Gojcic, Zan},
  journal = {Conference on Computer Vision and Pattern Recognition (CVPR)},
  year    = {2025}
}

@article{raysplats,
  title   = {{RaySplats: Ray Tracing Based Gaussian Splatting}},
  author  = {Byrski, Krzysztof and Mazur, Marcin and Tabor, Jacek and Dziarmaga, Tadeusz and Kadziolka, Marcin and Baran, Dawid and Spurek, Przemyslaw},
  journal = {arXiv preprint arXiv:2501.19196},
  year    = {2025}
}

@inproceedings{barron2022mip,
  title     = {{Mip-NeRF 360: Unbounded Anti-Aliased Neural Radiance Fields}},
  author    = {Barron, Jonathan T and Mildenhall, Ben and Verbin, Dor and Srinivasan, Pratul P and Hedman, Peter},
  booktitle = {{Proceedings of the IEEE/CVF Conference on Computer Vision and Pattern Recognition (CVPR)}},
  pages     = {5470--5479},
  year      = {2022}
}

\end{document}